\documentclass[aps,showpacs,preprintnumbers,amsmath,amssymb]{revtex4}

\usepackage{graphicx}
\usepackage{epstopdf}
\usepackage{color}
\usepackage{aasmacros}
\usepackage{amsmath,amssymb,amsfonts}

\begin{document}
    \baselineskip=0.8 cm
    \title{\bf Geometrically thick equilibrium tori around a Schwarzschild black hole in swirling universes}

    \author{Chengjia Chen$^{1}$, Qiyuan Pan$^{1,2}$\footnote{panqiyuan@hunnu.edu.cn}, and Jiliang Jing$^{1,2}$\footnote{jljing@hunnu.edu.cn} }

    \affiliation{$^1$ Department of Physics, Key Laboratory of Low Dimensional Quantum Structures and Quantum Control of Ministry of Education, Institute of Interdisciplinary Studies, and Synergetic Innovation Center for Quantum Effects and Applications, Hunan Normal University,  Changsha, Hunan 410081, China
    \\
    $^2$ Center for Gravitation and Cosmology, College of Physical Science and Technology, Yangzhou University, Yangzhou 225009, China}

    \begin{abstract}
    \baselineskip=0.6 cm
    \begin{center}
    {\bf Abstract}
    \end{center}

          We study geometrically thick non-self gravitating equilibrium tori orbiting a Schwarzschild black hole immersed in swirling universes. This solution is axially symmetric and non-asymptotically flat, and its north and south hemispheres spin in opposite directions. Due to repulsive effects arising from the swirl of the background spacetime, the equilibrium torus exists only in the case with the small swirling parameter. With the increase of the swirling parameter, the disk structure becomes small and the excretion of matter near the black hole becomes strong. Moreover, the odd $Z_2$ symmetry of spacetimes originating from the swirling parameter yields that the orientation of closed equipotential surfaces deviates away from the horizontal axis and the corresponding disk does not longer possess the symmetry with respect to the equatorial plane. These significant features could help to further understand the equilibrium tori and geometrically thick accretion disks around black holes in swirling universes.

    \end{abstract}

    \maketitle
    \newpage

\section{Introduction}

The real astrophysical black holes in the galaxies at the center of many galaxies are surrounded by an accretion disk. The matter flows in disks spiral into the central celestial body and drop off their initial angular momentums outwards together with releasing their gravitational potential energy into heat. The radiation emitted from accretion disks carries many characteristic information on black holes so that the corresponding observations and analysis can help to detect black holes and to examine theories of gravity in the strong field regime. The matter accretion near black holes is a highly complicated dynamic process and its full description can be completed only by resorting to high precise numerical simulations \cite{disc1,disc2,disc3,disc4,disc5,disc6}. In the past few decades, a simple and analytical model of geometrically thick and stationary tori orbiting black holes has been attracted a lot of attention. In this theoretical model, the matter motion in thick disks is simplified to be stationarily rotating and there is no actual accretion by the black hole \cite{smodel1,smodel2,smodel3,smodel4,smodel5,smodel6,smodel7,smodel8,smodel9}, and the self-gravity of the fluid body is also neglected. However, it is interesting that the stationarily rotating perfect fluid tori known as Polish doughnuts satisfy the relativistic Euler equation. Therefore, Polish doughnuts are always treated as a feasible initial condition to numerically simulate accretion flows near black holes. Moreover, this simplified model could capture key features of accretion disk configurations around black holes. Thus, the geometrically thick equilibrium tori without self-gravity  have been extensively studied in many black hole spacetimes in or beyond general relativity \cite{gr1,gr2,gr3,gr4,gr5,gr6,gr7,gr8,gr9,gr10,gr11,gr12,gr13,gr14,gr15}.

Recently, the geometrically thick equilibrium tori have been investigated in the spacetime of a static and spherically symmetric black hole in the frame of Born-Infeld teleparallel gravity, and it is found that there exists only a single torus and the teleparallel gravity parameter makes the size of the torus become small \cite{BI1}. For the parameterised Rezzolla-Zhidenko black hole,  there are standard single-torus and non-standard double equilibrium tori solutions \cite{RZ1,midu2}, and the transitions between them can occur through regulating the specific angular momentum of the fluid. The non-standard double equilibrium tori solutions are also found in the background of a dyonic black hole with quasi-topological electromagnetism \cite{zhou1}.
The magnetized equilibrium tori around Kerr black hole with scalar hair have been respectively studied within the constant
angular momentum model \cite{ksca1} and the nonconstant angular momentum model \cite{ksca2}. The features of geometrically thick equilibrium tori deviating from those in the usual Kerr case \cite{gr9} could be used to examine the no-hair hypothesis by combining with future astronomical observations. The stationarily and geometrically thick tori with the constant angular momentum have been studied in the background of a static black hole in $f(R)$-gravity with a Yukawa-like modification to the Newtonian potential \cite{fr1},  which shows that the equilibrium tori have the notable configurations differed from those in the usual static black holes in the general relativity. Moreover, the magnetized equilibrium tori around a Kerr black hole are further studied with the nonconstant specific angular momentum distribution model \cite{ksv1}. The configurations of the geometrically thick tori have been also studied in the background of a binary black hole system \cite{Binary1} and of compact object with a quadrupole moment \cite{qq1}.

We here focus on a Schwarzschild black hole solution in swirling universes \cite{bhswirl} satisfied the vacuum Einstein field equations. It is constructed analytically by exploiting the Ernst formalism \cite{ernst1,ernst2} with a transformation embedding a Schwarzschild seed spacetime into a rotating background. The rotating background can be interpreted as a gravitational whirlpool and its frame dragging yields that a static seed solution turns into a stationary metric \cite{bhswirl}. This black hole spacetime is axially symmetric and non-asymptotically flat, and its north and south hemispheres spin in opposite directions. Especially, there is no closed time-like curves in such spacetimes embedding in swirling universes, which means that it is less exotic than the G\"{o}del or the Taub-NUT spacetime \cite{bhswirl}. The Schwarzschild black hole solution in swirling universes has an event horizon at $r=2M$, which is the same as that of a usual Schwarzschild black hole and does not depend on the swirl of the background. However, the presence of the rotating background deforms the horizon
geometry so that it becomes more oblate, while the horizon area does not change. Moreover, the rotating background yields that this black hole also possesses an ergoregion, but its ergosurface is disconnected and consists of two non-compact branches above and below the black hole. The study of particles' geodesics \cite{Geode} shows that the geodesic equations can no longer be decoupled because the  black hole is immersed in the swirling universe and the spacetime is no longer of Petrov type D,  and  a small deviation of the swirling parameter from zero can change the qualitative features of the orbits significantly as compared to those in the Schwarzschild spacetime. Moreover, it is found that the swirling parameter drives the light rings outside the equatorial plane and the contour of the shadow becomes a tilted oblate shape \cite{shadows1}.
These studies shed new light on understanding the Schwarzschild black hole solution in swirling universes.
The main motivation of this work is to study the geometrically thick equilibrium tori around the Schwarzschild black hole solution in swirling universes and to see what new properties of the equilibrium tori in this case.

The work is organized as follows. In Sec. II, we briefly introduce the Schwarzschild black hole solution in swirling universes \cite{bhswirl} and Polish doughnut model of thick accretion disks \cite{smodel1,smodel2,smodel3,smodel4,smodel5,smodel6,smodel7,smodel8,smodel9}. In Sec. III, we investigate properties of equilibrium tori around the Schwarzschild black hole in the swirling background. Finally, we will include our conclusions in the last section.

\section{Schwarzschild black hole in swirling universes and Polish doughnut model of thick accretion disks}

Lets us first briefly review the Schwarzschild black hole solution in swirling universes \cite{bhswirl}, which is a stationary, axially symmetric and non-asymptotically flat black hole solution of the vacuum Einstein equations and can be obtained through an Ehlers transformation on the Schwarzschild solution.
The metric has a form as follows \cite{bhswirl}
\begin{equation}
\label{metric1}
{\rm d}s^2=F(r,\theta)\bigg(-f{\rm d}t^2+\frac{1}{f}{\rm d}r^2+ r^2{\rm d}\theta^2\bigg)+\frac{r^2\sin^2\theta}{F(r,\theta)}\bigg({\rm d}\phi+4jrf\cos\theta  {\rm d}t\bigg)^2,
\end{equation}
with
\begin{equation}
f=1-\frac{2M}{r},\quad\quad\quad\quad F(r,\theta)=1+j^2r^4\sin^4\theta.
\end{equation}
where $M$ is the black hole mass and $j$ is the parameter related to the spacetime rotation. The black hole's event horizon lies at $r=2M$ and the intrinsic singularity is located at $r=0$, which are the same as those of a Schwarzschild black hole. The presence of the swirling background deforms the horizon
geometry and  makes it more oblate, while the horizon area still remain $A=16\pi M^2$, which is independent of the swirling parameter. Moreover, due to the spacetime rotation, the black hole spacetime (\ref{metric1}) also possesses an ergoregion outside the event horizon $r=2M$, which is defined by $g_{tt}=0$. The ergoregions are located not only in the vicinity of the event horizon, but also close to the $z$-axis, for large values of $z$ \cite{bhswirl}.  Unlike in the Kerr black hole spacetime, the ergosurface in the swirling background (\ref{metric1}) is not a closed surface and two disconnected non-compact patches distribute above and below the black hole. With the increase of $j$, the  ergosurface gets closer to the horizon surface. The angular velocity $\Omega$ due to the frame dragging is  \cite{bhswirl,shadows1}
\begin{equation}
\Omega=\frac{d\phi}{dt}=-\frac{g_{t\phi}}{g_{\phi\phi}}=-4j(r-2M) \cos\theta,
\end{equation}
which means that the angular velocity $\Omega$ is in opposite directions in the two hemispheres.
In the spacetime (\ref{metric1}), since the
angular coordinate $\phi$ is periodic, an azimuthal curve $\gamma=\{ t=const; r=const; \theta= const\}$ is a closed curve with
the invariant length $s^2_{\gamma}=(2\pi)^2 g_{\phi\phi}$. Thus, there are no closed timelike curves since $g_{\phi\phi}$ is always non-negative, which means that
the Schwarzschild black hole in swirling universes is not plagued with geometric pathologies including conical singularities and causality violation originating from  closed timelike curves. When the parameter $j=0$, the metric (\ref{metric1}) recovers the Schwarzschild solution, and when $M=0$, it
reduces to the swirling universe.

Let us now to study geometrically thick equilibrium tori around the  Schwarzschild black hole in swirling universes (\ref{metric1}). With the test-fluid approximation \cite{smodel1,smodel2,smodel3,smodel4,smodel5,smodel6,smodel7,smodel8,smodel9}, the accretion flow in the disk is assumed as a barotropic perfect fluid with the positive pressure and the disk self-gravity is
negligible. The fluid is assumed to be axisymmetric and stationary in the spacetime and the corresponding physical variables
depend on only the coordinated $r$ and $\theta$. Moreover, the rotation of the perfect fluid is restricted to be in the
azimuthal direction. Under these assumptions, the four-velocity of the perfect fluid can be expressed as $u^{\mu}=(u^t,0,0,u^{\phi})$
\cite{smodel1,smodel2,smodel3,smodel4,smodel5,smodel6,smodel7,smodel8,smodel9},
and the corresponding stress-energy tensor is
\begin{eqnarray}
T_{\mu\nu}=w u_\mu u_\nu + p g_{\mu\nu},
\end{eqnarray}
where $w$ and $p$ are the enthalpy density and the pressure of the fluid, respectively. Making use of the conservation for the perfect fluid $\nabla_\mu T_\nu^\mu=0$, one can obtain \cite{smodel1,smodel2,smodel3,smodel4,smodel5,smodel6,smodel7,smodel8,smodel9}
\begin{equation}
-\nabla_{i} \ln |u_t|+\frac{\Omega \nabla_{i} l}{1-l \Omega}=\frac{1}{w} \nabla_{i} p, \quad\quad\quad i=r, \;\theta,\label{eq10}
\end{equation}
where $l$ is the specific angular momentum of the fluid particle, and $u_t$ is the redshift factor and its form in the background spacetime (\ref{metric1}) can be expressed as
\begin{equation}
u_t=\sqrt{\frac{g^2_{t \phi}-g_{t t} g_{\phi \phi}}{l^2 g_{t t}+2g_{t\phi}l+g_{\phi \phi}}}.\label{eq9}
\end{equation}
The quantity $\Omega$ is the angular velocity of the fluid particle
\begin{equation}
 \Omega=\frac{d \phi}{d t}=-\frac{g_{t\phi}+lg_{\phi\phi}}{g_{\phi\phi}+lg_{t\phi}}.\label{Omegal}
\end{equation}
According to the von Zeipel theorem \cite{smodel1,smodel2,smodel3,smodel4,smodel5,smodel6,smodel7,smodel8,smodel9}, one has $\Omega=\Omega(l)$ and
can further obtain
\begin{equation}
W-W_{in}=\ln \left|u_t\right|-\ln \left|\left(u_t\right)_{\mathrm{in}}\right|-\int_{l_{\mathrm{in}}}^{l}\left(\frac{\Omega}{1-\Omega l'}\right) d l',
\end{equation}
where the subscript ``in" denotes that the quantity is evaluated at the inner edge of the disk. The potential $W$ is given by
\begin{equation}
W-W_{in}=\int_0^{p}\frac{dp}{w},\label{wwdp}
\end{equation}
which determines the shapes of equipotential surfaces in the disk. Assuming that a barotropic equation of state takes a form $P=K w^{\kappa}$  with two constants $K$ and $\kappa$ \cite{midu1}, from Eq. (\ref{wwdp}), one can easily obtain
\begin{equation}
W-W_{in} + \frac{\kappa}{\kappa-1}\frac{p}{w}=0.\label{wwinp}
\end{equation}
Thus, once the potenial $W$ and the specific enthalpy $h=w/\rho$ are given, one can get the enthalpy density $w$ and the rest-mass density $\rho$ distribution in the disk. For simplicity, one can set  $h=1$ as in Ref. \cite{midu1}. Moreover, we here focus on only a simple case where the fluid has a constant specific angular momentum $l$, which means the fluid angular velocity $\Omega$ in Eq. (\ref{Omegal}) becomes a function only related to the spacetime metric.  In this simple model, the potential $W$ can be further simplified as $W=\ln|u_t|$. Such stationarily rotating perfect-fluid tori are known as ``Polish doughnuts".

\section{Geometrically thick equilibrium tori around the Schwarzschild black hole in swirling universes}

\begin{figure}
   \includegraphics[width=6.5cm]{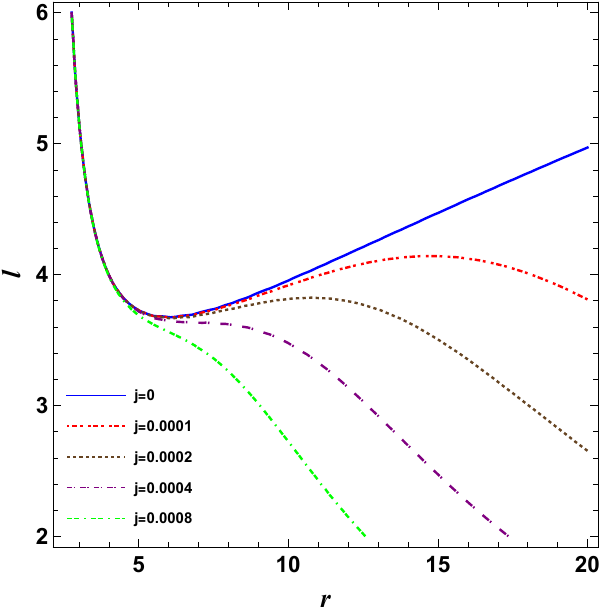}\includegraphics[width=6.5cm]{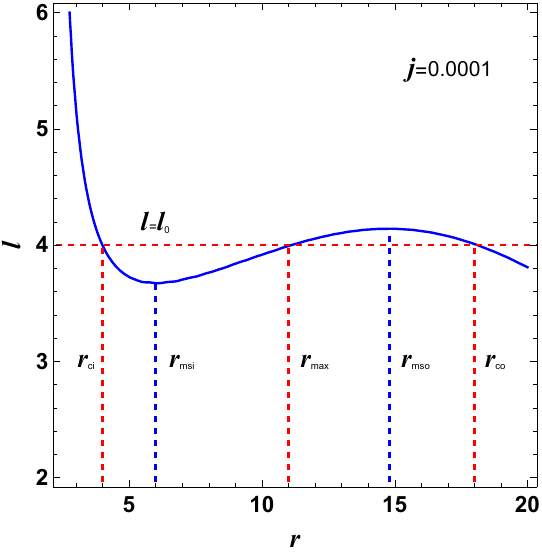}
   \caption{Effects of the swirling parameter $j$ on the specific angular momentum $l$ for the particles moving along circular orbits in the spacetime of a Schwarzschild black hole in swirling universes. Here we set $M=1$.}
   \label{lkms}
   \end{figure}

Thick equilibrium tori are allowed only in the spacetimes admitting closed equipotential $W$ surfaces. The outermost closed surface crosses itself in the cusp(s). These cusps enable the outflow of matter from the torus because of the breaking of hydrostatic equilibrium. In the usual Schwarzschild spacetime, there is only an inner cusp enabling an accretion onto the
central black hole. In the spacetimes with repulsive parameters including cosmological constant \cite{gr4,gr5,gr6}, there also exists the outer cusp on the equipotential surfaces at which the gravitational attraction is just balanced
by the corresponding repulsion. The outer cusp enables the excretion where matter outflows from the torus into the outer space.
In the case of a Schwarzschild black hole in swirling universes, the potential $W$ has a form
\begin{equation}
W(r,\theta)=\frac{1}{2}\ln\frac{(r-2M)r^2\sin^2\theta(1+j^2r^4\sin^4\theta)}{r^3[1+4jl(r-2M)\cos\theta]^2\sin^2\theta-(r-2M)(1+j^2r^4\sin^4\theta)^2l^2}.\label{wform}
\end{equation}
The features of the equipotential surfaces of fluid in the Schwarzschild black hole spacetime with the swirling parameter $j$ are fully determined by the number of the local extrema of the function $W(r, \theta)$ and their mutual behavior. The local extrema of $W(r, \theta)$ can be obtained by the conditions
 \begin{eqnarray}\label{wextr}
  \frac{\partial W(r, \theta)}{\partial r}=0,\quad\quad\quad \frac{\partial W(r, \theta)}{\partial \theta}=0.
 \end{eqnarray}
Unlike in the usual Schwarzschild case, the local extrema of $W(r, \theta)$ is not located in the equatorial plane except the case $j=0$, which is understandable because the spacetime (\ref{metric1}) possesses only the odd $Z_2$ (north-south) symmetry. Moreover, the potential $W$ also has the symmetry under the transformation $(j,\theta)\leftrightarrow (-j,\;\pi-\theta)$, so we here consider only the case $j\geq 0$ without loss of generality. The number  of the local extrema of the function $W(r, \theta)$ is determined by the parameter $j$ and the specific angular momentum $l$. Solving numerically Eq. (\ref{wextr}), we can obtain the specific angular momentum $l$ for the particles moving along circular orbits in the spacetime (\ref{metric1}).
Fig. \ref{lkms} exhibits the effects of the swirling parameter $j$ on  the changes of the specific angular momentum $l$ with circular orbit radii $r$. As $j=0$, one can find that there is only a minimum $l_{\rm ms}=\frac{3}{2}\sqrt{6}$ at $r=r_{\rm ms}$, which recovers the results obtained in the usual Schwarzschild case. With the increase of $j$, we find that the function $l$ has two local extrema, i.e., the minimum $l_{\rm msi}$ and the maximum $l_{\rm mso}$, which respectively correspond to the inner and outer marginally stable circular geodesics. The rising (descending)  part of $l$ corresponds to stable (unstable) circular geodesics. As the swirling parameter increases up to $j>j_{c1}=0.000378$, the function $l$ becomes monotonically decreasing, which means that there exist only unstable circular geodesics in this case. Therefore, as in the Schwarzschild-de Sitter spacetime, a toroidal configuration of finite size can be obtained only for values of the specific angular
momentum $l_0$ that satisfy
\begin{eqnarray}\label{lmsktt}
 l_{\rm msi} <l_0<l_{\rm mso},
 \end{eqnarray}
and there is no the toroidal configuration around the black hole (\ref{metric1}) as the swirling parameter $j>0.000378$. Therefore, the swirling parameter $j$ plays a crucial role in the existence of the disk solutions. The slight increase of $j$ could yield that the possibility of having solutions decreases dramatically. This means that this disk's model exists only for the relatively small swirling parameter.

\begin{figure}
   \includegraphics[width=7.5cm]{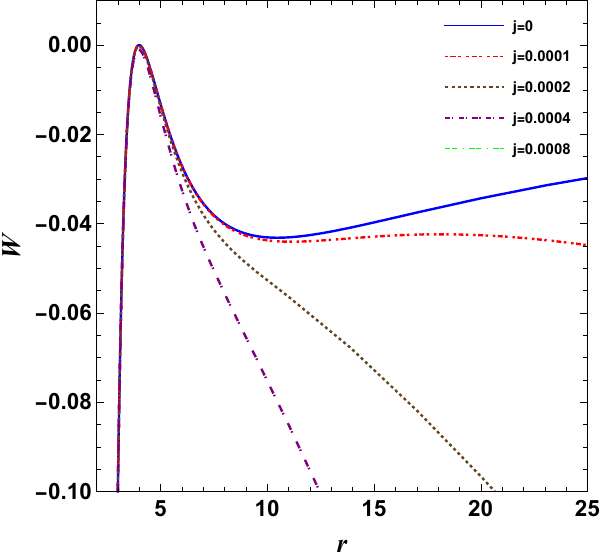}\includegraphics[width=7.5cm]{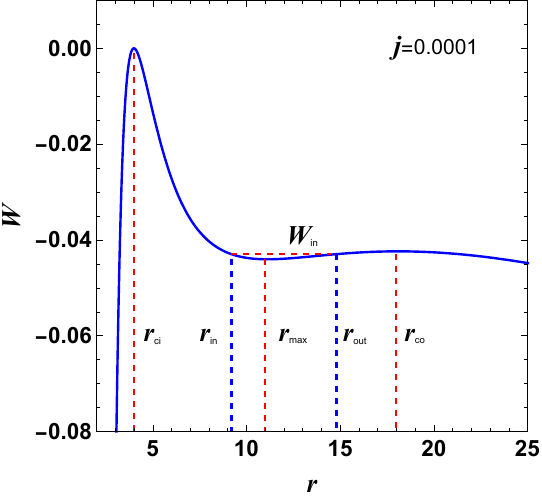}
   \caption{Effects of the swirling parameter $j$ on the effective potential $W$ for the particles moving along circular orbits in the spacetime of a Schwarzschild black hole in swirling universes for fixed specific angular momentum $l=4$. Here we set $M=1$.}
   \label{wkms}
\end{figure}

\begin{figure}
   \includegraphics[width=5cm]{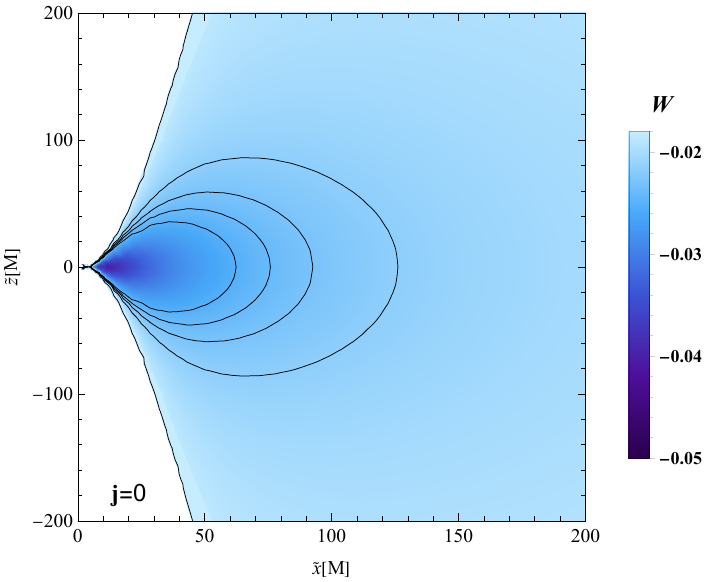}\includegraphics[width=5cm]{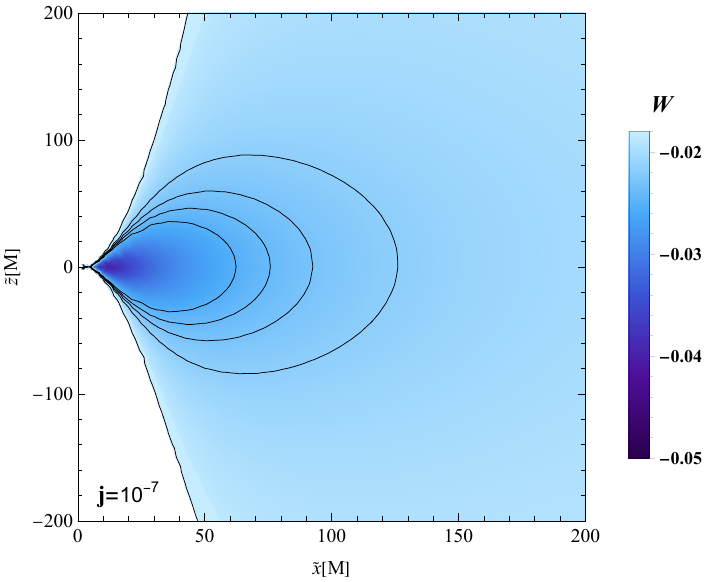}\includegraphics[width=5cm]{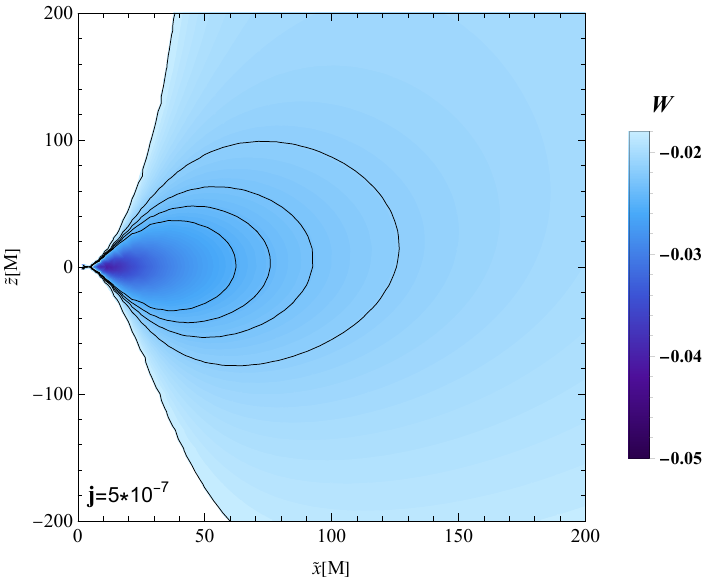}\\
   \includegraphics[width=5cm]{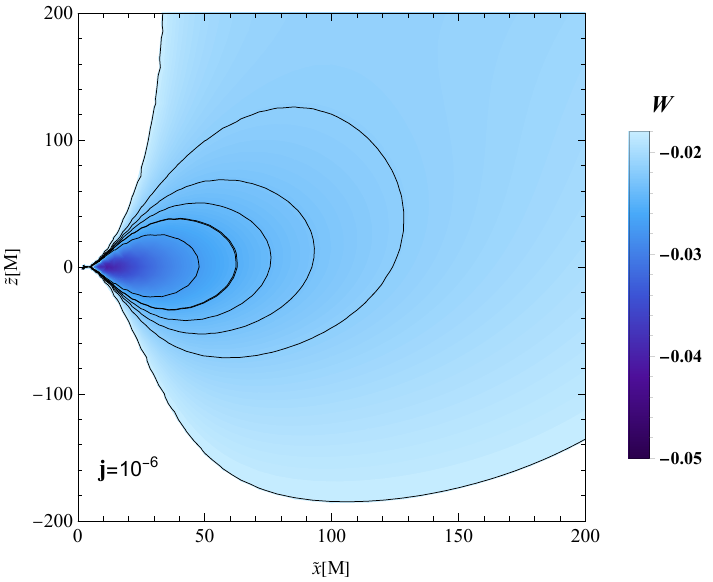}\includegraphics[width=5cm]{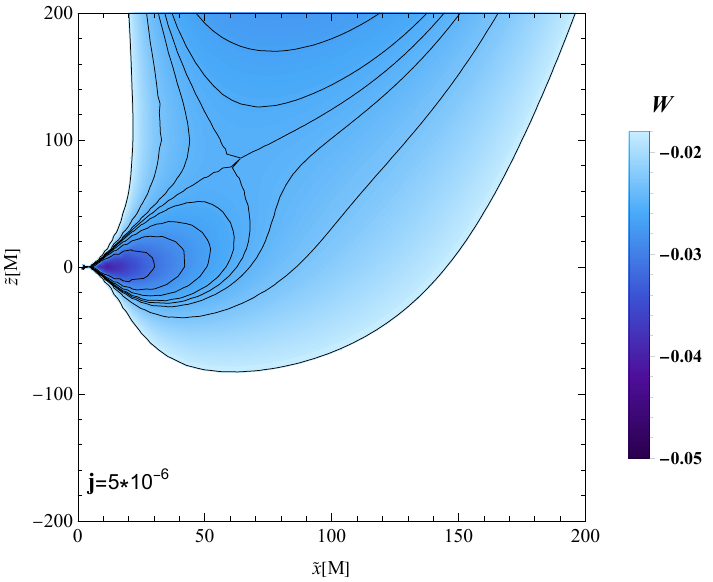}\includegraphics[width=5cm]{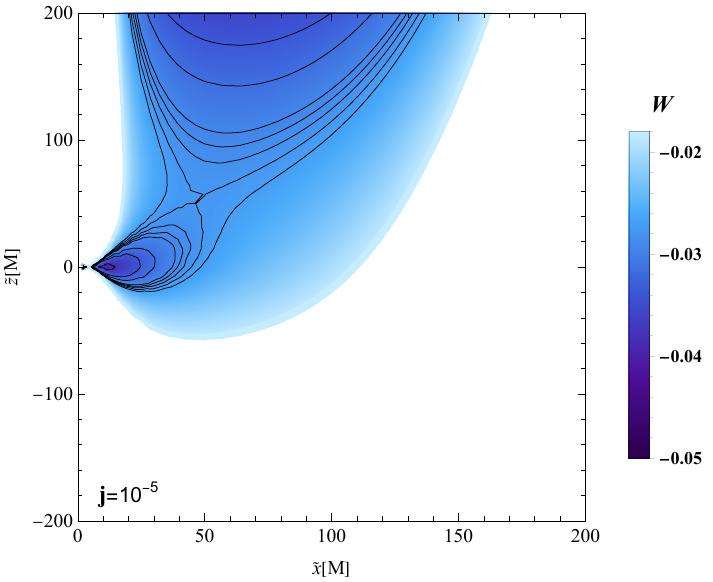}\\
   \includegraphics[width=5cm]{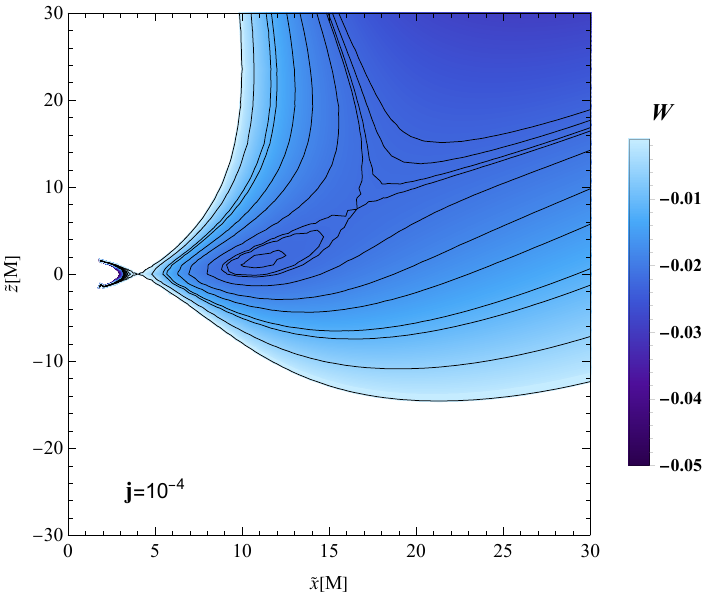}\includegraphics[width=5cm]{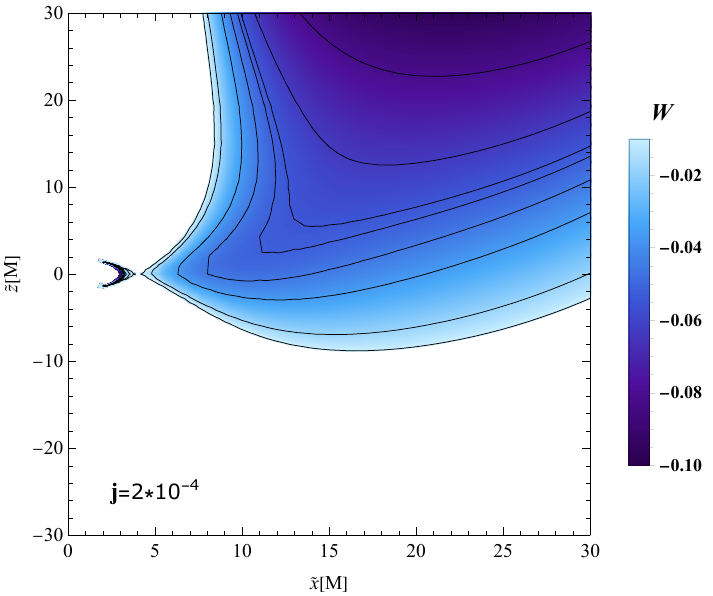}\includegraphics[width=5cm]{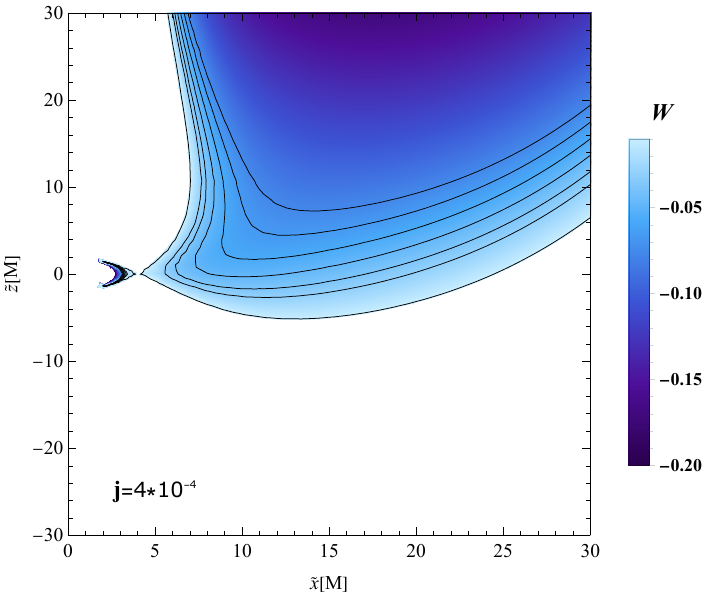}
   \caption{Equipotential surfaces of the effective potential $W$ shown with Cartesian coordinates $\tilde{x}=r\sin\theta \cos\phi$ and $\tilde{z}=r\cos\theta$ for different swirling parameters $j$. Here we set $M=1$ and $l=4$.}
   \label{wrjyes}
   \end{figure}

Fig. \ref{wkms} presents the effects of $j$ on numbers of the local extrema of $W(r, \theta)$ for $l=4$ as the fluid particles move along circular orbits. In the non-zero $j$ cases, we find that $W(r, \theta)$ has three local extrema as $j<j_{c2}=0.000132$. Two local maximum points respectively correspond to the positions of the inner cusp $ r_{\rm ci}$ and the outer cusp $ r_{\rm co}$, where the equipotential surface has a self-crossing point in the $(r, \theta)$ plane.
The outer cusp disappears in the Schwarzschild spacetime but emerges in the de Sitter spacetimes with the non-zero cosmological constant. The presence of the outer cusp represents that the swirling parameter yields some repulsive effects similar to the cosmological constant. Fig. \ref{wkms} shows the casse $W(r_{\rm ci})>W(r_{\rm co})$, which means that the fluid particles with $W(r_{\rm co})<W<W(r_{\rm ci})$ moving along the circular orbit can be excreted from the torus to the spatial infinity rather than be accreted into the black hole. This does not appear in the usual Schwarzschild black hole case.
 The local minimum point  is located at the ``centre" of the torus $r_{\rm max}$ where the internal pressure has its local
maximum in the torus. As $j\geq j_{c2}=0.000132$, we find that the function $W$ has no local extremum.
These are also shown in Fig. \ref{wrjyes}, in which the equipotential surfaces of the effective potential $W$ are given in Cartesian coordinates $\tilde{x}[M]$ and $\tilde{z}[M]$.
As $j=0$, the equipotential surfaces possess the equatorial symmetry. However, the presence of the swirling parameter $j$ leads to that the equipotential surfaces do not have the equatorial symmetry and the closed equipotential surfaces orient away from the horizontal axis. This can be understood by the asymmetry with respect to the equatorial plane originating from the swirl of the background spacetime. The disk deviation from the horizontal axis noticeably increases with the parameter $j$. The orientation of equipotential surfaces deviating from the horizontal axis also appears in the case of accelerating black holes \cite{gr7}, which can be regarded as a common feature of equipotential surfaces in the spacetimes only with the odd $Z_2$ symmetry.
Moreover, the distance between the center and outer cusp points changes as a monotonic decreasing function of $j$. Thus, the increase of $j$ results in the smaller disk structure. As $j$ increases up to a threshold that the center
and the cusp point overlaps, there is no closed equipotential surfaces and then the disk structure vanishes. This means that the larger disk structure is only for the small swirl of the spacetime.

\begin{figure}
\includegraphics[width=5cm]{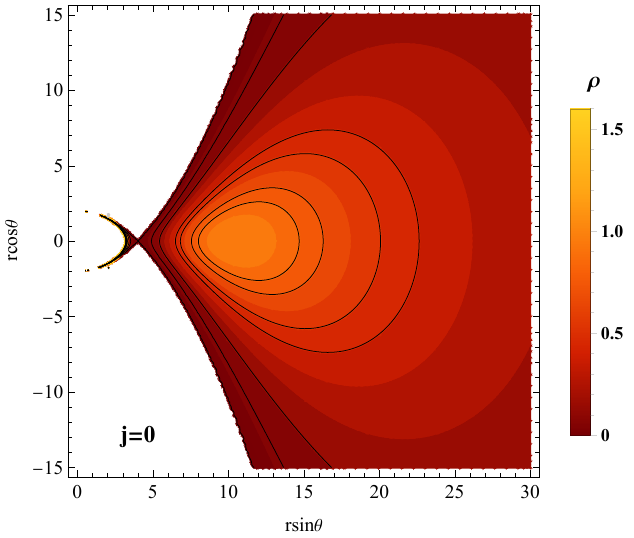}\includegraphics[width=5cm]{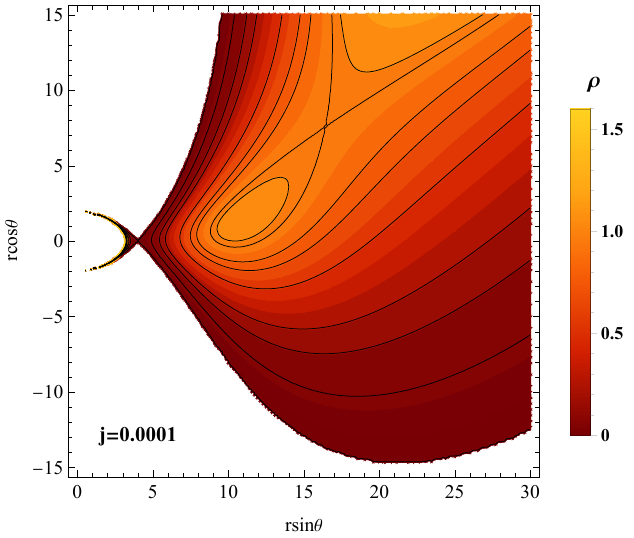}\includegraphics[width=5cm]{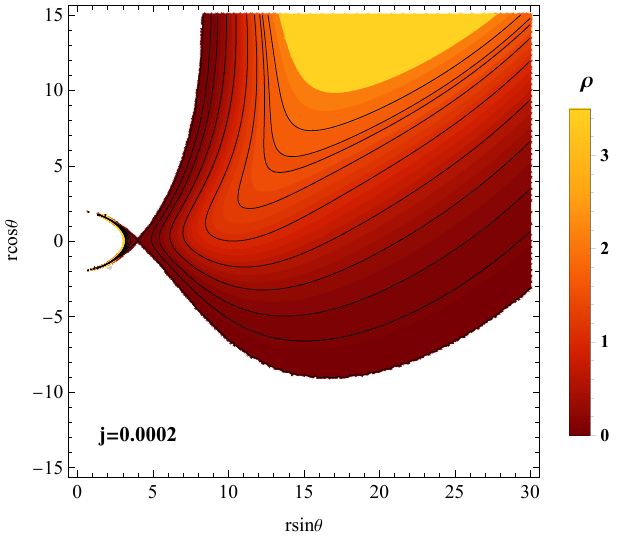}
   \caption{Contour map of the rest-mass density of the disk for different swirling parameters $j$. Here we set $M=1$ and $l=4$.}
   \label{massdensity}
\end{figure}

In Fig. \ref{massdensity}, we also present the distributions of the rest-mass density in equilibrium tori by solving Eq. (\ref{wwinp}) with the equation of state $p=K w^{\kappa}$ through setting $\kappa=4/3$ and $w_{r_{\rm max}}=1$. The coefficient $K$ is calculated by $K=W_{r_{\rm max}}-W_{\rm in}$, where $W_{r_{max}}$ and $W_{\rm in}$ are respectively  the values of $W$ at the centre $r=r_{\rm max}$ of the equilibrium torus and at the inner cusp. Comparing Fig. \ref{massdensity} with Fig. \ref{wrjyes}, it is easy to observe that the distributions of mass density in equilibrium tori are similar to the distribution of the effective potential $W$. With the increase of $j$, one can find that the excretion of matter near the black hole becomes strong. These results could help to further understand equilibrium tori and geometrically thick accretion disks around the Schwarzschild black hole in swirling universes.

\section{Summary}

We have studied geometrically thick non-self gravitating equilibrium tori orbiting a Schwarzschild black hole in swirling universes. This solution is non-asymptotically flat, and its north and south hemispheres spin in opposite directions. Due to repulsive effects arising from the swirl of the background spacetime, the equilibrium torus exists only in the case with the small swirling parameter. In the case of existing equilibrium tori, there are the inner cusp $ r_{\rm ci}$ and the outer cusp $ r_{\rm co}$ for equipotential surfaces and one has $W(r_{\rm ci})>W(r_{\rm co})$, meaning that the fluid particles with $W(r_{\rm co})<W<W(r_{\rm ci})$ moving along the circular orbit can be excreted from the torus to the spatial infinity rather than  be accreted into the black hole, which does not appear in the usual Schwarzschild black hole case. With the increase of the swirling parameter $j$, one can find that the disk structure becomes small and the excretion of matter near the black hole becomes more strong. Moreover, the odd $Z_2$ symmetry of spacetimes originating from the swirling parameter yields that the orientation of closed equipotential surfaces deviates away from the horizontal axis so that the disk does not longer possess the symmetry with respect to the equatorial plane. Thus, the swirling parameter $j$ plays a crucial role in the existence and behavior of the equilibrium tori around the black hole. These results could help to understand the equilibrium tori and geometrically thick accretion disks around black holes in swirling universes, and to further distinguish such black holes from others by analyzing the observing features of accretion disks.

\begin{acknowledgments}

This work was supported by the National Key Research and Development Program of China (Grant No. 2020YFC2201400) and National Natural Science Foundation of China (Grant Nos. 12275079 and 12035005).

\end{acknowledgments}

    \end{document}